%% file: main.tex
\documentclass{style/technical_report}

\usepackage{graphicx}
\usepackage{xcolor}
\usepackage{listings}

\usepackage{textcomp}
\usepackage[ruled,lined]{algorithm2e}
\usepackage{cite}
\usepackage{array}

\input{macros/macros.tex}

\begin{document}

\title{Record-replay debugging for the SCOOP concurrency model}

\author{Benjamin Morandi, Sebastian Nanz, and Bertrand Meyer
\and
{\small
\begin{tabular}{c}
Chair of Software Engineering, ETH Zurich, Switzerland \\
\texttt{firstname.lastname@inf.ethz.ch} \\ 
\texttt{http://se.inf.ethz.ch/}
\end{tabular}}
}

\maketitle

\begin{abstract}
To support developers in writing reliable and efficient concurrent programs, novel concurrent programming abstractions have been proposed in recent years. Programming with such abstractions requires new analysis tools because the execution semantics often differs considerably from established models. We present a record-replay technique for programs written in SCOOP, an object-oriented programming model for concurrency. The resulting tool enables developers to reproduce the nondeterministic execution of a concurrent program, a necessary prerequisite for debugging and testing.
\end{abstract}

\input{introduction}
\input{related_work}
\input{SCOOP}
\input{record_replay}
\input{conclusion}

\bibliography{bibliography}

\end{document}

%% file: macros/macros.tex
\newcommand{\sectionreference}[1]{Section~\ref{#1}}
\newcommand{\figurereference}[1]{Figure~\ref{#1}}

\newcommand{\algorithmreference}[1]{Algorithm~\ref{#1}}
\SetKw{triggerinstruction}{trigger}
\SetKw{waitinstruction}{wait}
\SetKwData{undefinedvalue}{undef}

\SetKwBlock{multilineeventhandlerstart}{upon event}{}
\SetKw{singlelineeventhandlerstart}{upon event}
\SetKw{eventhandlermiddle}{do}
\SetKw{eventhandlerconditionstart}{such that}
\newcommand{\eventhandler}[3]{\multilineeventhandlerstart(#1 \eventhandlermiddle{} #2){#3}}

\SetKwBlock{multilineconditionhandlerstart}{upon}{}
\SetKw{singlelineconditionhandlerstart}{upon}
\SetKw{conditionhandlermiddle}{do}

%% file: introduction.tex
\section{Introduction}
\label{sec:introduction}
Avoiding concurrency-specific errors such as data races and deadlocks is still the responsibility of developers in most languages that provide synchronization through concurrency libraries. To avoid the problems of the library approach, a number of languages have been proposed that fully integrate synchronization mechanisms. SCOOP (Simple Concurrent Object-Oriented Programming) \cite{meyer:1997:oosc,nienaltowski:2007:SCOOP}, an object-oriented programming model for concurrency, is one of them.

The main idea of SCOOP is to simplify the writing of correct concurrent programs, by allowing developers to use familiar concepts from object-oriented programming, but protecting them from common concurrency errors such as data races. Empirical evidence supports the claim that SCOOP indeed simplifies reasoning about concurrent programs as opposed to more established models~\cite{nanz-torshizi-pedroni-meyer:2011:usability_of_SCOOP}.

The complex interactions between concurrent components make it difficult to analyze the behavior of typical concurrent programs. Effective use of a programming model therefore requires tools to help developers analyze and improve programs. Static analysis of models, e.g.,\ \cite{ostroff-torshizi-huang-schoeller:2008:formal_semantics_for_SCOOP,brooke-paige-jacob:2007:formal_semantics_for_SCOOP,west-nanz-meyer:2010:deadlock_prevention_for_SCOOP,nanz-nielson-nielson:2008:modal_abstractions_of_concurrent_behavior}, can establish some degree of functional correctness. However, they fail to explain why a particular execution does not terminate. Once a problem has been identified, it may be difficult to reproduce it because the problem might manifest itself only under some particular interleavings. Worse, the act of debugging itself might make it go away because of changes in the interleaving caused by the observation instructions. The term \emph{Heisenbug} is sometimes used to denote this phenomenon. Addressing these issues requires adapting record-replay techniques to the context of concurrent, non-deterministic execution. \sectionreference{sec:related-work} surveys existing tools that address this goal. They are not appropriate, however, for the semantics of SCOOP.

We present a SCOOP adaptation of Choi and Srinivasan's \cite{choi-srinivasan:1998:replay_for_concurrent_programs} record-replay technique for Java threads. The resulting tool has been integrated into the EVE~\cite{eth:2011:EVE} development environment, which we extended with support for SCOOP.
We found that the SCOOP model provides abstractions that can be leveraged by the technique: SCOOP's synchronization mechanism provides abstractions which are coarse-grained enough to limit state space explosion and thus keep execution records small.

This article is structured as follows. \sectionreference{sec:related-work} provides an overview of related work. \sectionreference{sec:SCOOP} gives an overview of the SCOOP model. \sectionreference{sec:recordreplay} presents the adapted record-replay technique. \sectionreference{sec:conclusion} concludes with an outlook on future work.

%% file: related_work.tex
\section{Related work}\label{sec:related-work}
The main problem of debugging concurrent programs is to make concurrent executions repeatable; a number of
approaches to address this problem have emerged. The approach of
Pan and Linton~\cite{pan-linton:1989:replay_for_concurrent_programs} logs all data read from
shared memory locations. To replay, it simulates the events from the log. While this
approach has the advantage of allowing immediate reverse execution
of a program (backstepping), its main drawback is the prohibitively
large amount of data generated during execution, as acknowledged
by~\cite{pan-linton:1989:replay_for_concurrent_programs}.

Most approaches have, as a consequence, focused on recording only the
order of events, not the data; in a second step this information is
used to replay the execution. The predominant approaches can be
classified according to the type of information recorded: either only
coarse-grained information such as object accesses and synchronization
events~\cite{leblanc-mellor-crummey:1987:replay_for_concurrent_programs,tai-carver-obaid:1991:replay_for_concurrent_programs}
or every shared-memory access~\cite{netzer:1993:replay_for_concurrent_programs}.  LeBlanc and
Mellor-Crummey~\cite{leblanc-mellor-crummey:1987:replay_for_concurrent_programs} describe a
method termed \emph{Instant Replay} that records the order of accesses
to shared objects during a monitoring phase by assigning version
numbers to objects and recording for each process which object
versions have been accessed. Through this recording of object accesses,
it can be ensured during replay that processes access objects of the
same version numbers as during monitoring, thus reproducing the
execution and the object values. Tai et
al.~\cite{tai-carver-obaid:1991:replay_for_concurrent_programs} consider programs where all
shared objects are protected by synchronization mechanisms. They record
the order of the synchronization operations. During replay, the execution can thus
be recreated under the assumption that a program is free of data races.
Netzer~\cite{netzer:1993:replay_for_concurrent_programs} proposes monitoring every
shared-memory access so that data race-freedom no longer needs to be
assumed. The technique is optimized with regard to the amount of
information needed to reproduce an execution; it performs a transitive reduction
of the dependencies between shared-memory accesses and only records the optimal ordering, thus significantly reducing
the size of the trace log. A drawback of the approach consists, however,
in the large amount of runtime overhead, as pointed out
by~\cite{russinovich-cogswell:1996:replay_for_concurrent_programs}.

Bacon and Goldstein~\cite{bacon-goldstein:1991:replay_for_concurrent_programs} present a
hardware-assisted scheme for deterministic replay. In contrast to
the software-based methods, the scheme succeeds in avoiding the
complications of the probe effect.  Xu et
al.~\cite{xu-bodik-hill:2003:replay_for_concurrent_programs} develop this approach further using a variant of the transitive reduction~\cite{netzer:1993:replay_for_concurrent_programs} to minimize log size.

Instead of relying on a log of application events, as the previously discussed approaches usually do, Russinovich and
Cogswell~\cite{russinovich-cogswell:1996:replay_for_concurrent_programs} recreate program
executions by logging thread switches caused by the system scheduler. They modify the operating system to generate a log that can recreate the thread
switches upon replay. Choi and
Srinivasan~\cite{choi-srinivasan:1998:replay_for_concurrent_programs} further improve
this approach by logging \emph{logical thread schedules} representing
equivalence classes of physical thread schedules with respect to the
ordering of shared-memory access events. Our approach for
record-replay is based on logical thread schedules and adapts the idea in
the context of SCOOP.


%% file: SCOOP.tex
\section{SCOOP}\label{sec:SCOOP}
This section gives an overview of SCOOP. The starting idea of SCOOP is that every object is associated for its lifetime with a processor, called its \emph{handler}. A \emph{processor} is an autonomous thread of control capable of executing actions on objects. An object's class describes the possible actions as \emph{features}. A processor can be a CPU, but it can also be implemented in software, for example as a process or as a thread; any mechanism that can execute instructions sequentially is suitable as a processor.

A variable \lstinline[language=SCOOP]!x! belonging to a processor can point to an object with the same handler (\emph{non-separate object}), or to an object on another processor (\emph{separate object}). In the first case, a \emph{feature call} \lstinline[language=SCOOP]!x.f! is \emph{non-separate}: the handler of \lstinline[language=SCOOP]!x! executes the feature synchronously. In this context, \lstinline[language=SCOOP]!x! is called the \emph{target} of the feature call. In the second case, the feature call is \emph{separate}: the handler of \lstinline[language=SCOOP]!x!, i.e., the \emph{supplier}, executes the call asynchronously on behalf of the requester, i.e., the \emph{client}. The possibility of asynchronous calls is the main source of concurrent execution. The asynchronous nature of separate feature calls implies a distinction between a feature call and a \emph{feature application}: the client logs the call with the supplier (feature call) and moves on; only at some later time will the supplier actually execute the body (feature application).

The producer-consumer problem serves as a simple illustration of these ideas. A root class defines the entities \lstinline[language=SCOOP]!producer!, \lstinline[language=SCOOP]!consumer!, and \lstinline[language=SCOOP]!buffer!. Assume that each object is handled by its own processor. One can then simplify the discussion using a single name to refer both to the object and its handler. For example, one can use ``producer'' to refer both to the producer object and its handler.
\begin{lstlisting}[mathescape=true,language=SCOOP]
producer: separate PRODUCER
consumer: separate CONSUMER
buffer: separate BUFFER [INTEGER]
\end{lstlisting}
%
The keyword \lstinline[language=SCOOP]!separate! specifies that the referenced objects may be handled by a processor different from the current one. A \emph{creation instruction} on a separate entity such as \lstinline[language=SCOOP]!producer! will create an object on another processor; by default the instruction also creates that processor.

Both the producer and the consumer access an unbounded buffer in feature calls such as \lstinline[language=SCOOP]!buffer.put (n)! and \lstinline[language=SCOOP]!buffer.item!. To ensure exclusive access, the consumer must lock the buffer before accessing it. Such locking requirements of a feature must be expressed in the formal argument list: any target of separate type within the feature must occur as a formal argument; the arguments' handlers are locked for the duration of the feature execution, thus preventing data races. Such targets are called \emph{controlled}. For instance, in \lstinline[language=SCOOP]!consume!, \lstinline[language=SCOOP]!buffer! is a formal argument; the consumer has exclusive access to the buffer while executing \lstinline[language=SCOOP]!consume!.

Condition synchronization relies on preconditions (after the \lstinline[language=SCOOP]!require! keyword) to express wait conditions. Any precondition of the form \lstinline[language=SCOOP]!x.some_condition! makes the execution of the feature wait until the condition is true. For example, the precondition of \lstinline[language=SCOOP]!consume! delays the execution until the buffer is not empty. As the buffer is unbounded, the corresponding producer feature does not need a wait condition.
\begin{lstlisting}[language=SCOOP]
consume (buffer: separate BUFFER [INTEGER])
		-- Consume an item from the buffer.
	require not (buffer.count = 0)
	local
		consumed_item: INTEGER
	do
		consumed_item := buffer.item
	end
\end{lstlisting}
The runtime system ensures that the result of the call \lstinline[language=SCOOP]!buffer.item! is properly assigned to the entity \lstinline[language=SCOOP]!consumed_item! using a mechanism called \emph{wait by necessity}: while the consumer usually does not have to wait for an asynchronous call to finish, it will do so if it needs the result.

The SCOOP concepts require runtime support.
The following description is abstract; actual implementations may differ.
Each processor maintains a \emph{request queue} of requests resulting from feature calls on other processors. A non-separate feature call can be processed right away without going through the request queue; the processor creates a \emph{non-separate feature request} for itself and processes it right away using its call stack. When the client executes a separate feature call, it enqueues a \emph{separate feature request} to the request queue of the supplier's handler. The supplier will process the feature requests in the order of queuing.

The runtime system includes a \emph{scheduler}, which serves as an arbiter between processors. When a processor is ready to process a feature request in its request queue, it will only be able to proceed after the request is satisfiable. In a \emph{synchronization step}, the processor tries to obtain the locks on the arguments' handlers in a way that the precondition holds. For this purpose, the processor sends a \emph{locking request} to the scheduler, which stores the request in a queue and schedules satisfiable requests for application. Once the scheduler satisfies the request, the processor starts an \emph{execution step}.

Whenever a processor is ready to let go of the obtained locks, i.e., at the end of its current feature application, it issues an unlock request to each locked processor. Each locked processor will unlock itself as soon as it processed all previous feature requests. In the example, the producer issues an unlock request to the buffer after it issued a feature request for \lstinline[language=SCOOP]!put!.


%% file: record_replay.tex
\section{Record-replay}\label{sec:recordreplay}
This section presents a record-replay technique for SCOOP programs. The technique is an adaptation of Choi and
Srinivasan's~\cite{choi-srinivasan:1998:replay_for_concurrent_programs} approach, developed for
Java multithreading. Their notion
of logical thread schedules helps keep the size of the log file
small. \sectionreference{sec:logical-schedules} presents the SCOOP-adaptation of logical
thread schedules, called logical processor schedules. \sectionreference{sec:recording logical processor schedules} and \sectionreference{sec:replaying logical processor schedules} show how the SCOOP runtime records and replays them.

\subsection{Logical processor schedules}\label{sec:logical-schedules}
As demonstrated in \sectionreference{sec:related-work}, a number of effective approaches to the problem of deterministic replay of multithreaded programs exist. For executions on uniprocessor systems, the approach of Russinovich and Cogswell~\cite{russinovich-cogswell:1996:replay_for_concurrent_programs} has been shown to outperform techniques that try to record how threads interact. They propose to log thread scheduler information and to enforce the same schedule when a run is replayed.
This approach also works well in our case. To minimize the overhead from capturing
\emph{physical processor schedules} -- the equivalent of physical thread schedules in the case of
SCOOP -- we adapt the notion of \emph{logical thread schedules} from~\cite{choi-srinivasan:1998:replay_for_concurrent_programs}. This section describes this adaptation.

Consider a share market application with investors, markets, issuers, and shares. The markets and the investors are handled by different processors. Listing~\ref{lst:investor class} shows the class for the investors. Each investor has a feature to buy a share. To execute it, the investor must wait for the lock on the market and for the precondition to be satisfied.
\begin{lstlisting}[caption=Investor class, label=lst:investor class, language=SCOOP]
class INVESTOR feature
	id: INTEGER
	
	buy (market: separate MARKET; issuer_id: INTEGER)
			-- Buy a share of the issuer on the market.
		require
			market.can_buy (id, issuer_id)
		do
			market.buy (id, issuer_id)
		end
end
\end{lstlisting}
The following feature initiates a transaction that involves two investors and one market with shares from two issuers:
\begin{lstlisting}[language=SCOOP]
do_transaction (first_investor, second_investor: separate INVESTOR; base_issuer_id: INTEGER)
		-- Make the two investors buy two shares from two consecutive issuers on the market.
	local
		next_issuer_id: INTEGER	
	do
		first_investor.buy (market, base_issuer_id)
		next_issuer_id := base_issuer_id + 1
		second_investor.buy (market, next_issuer_id)
	end
\end{lstlisting}

\begin{figure}[ht]
  \centering
  \includegraphics[width=0.95 \textwidth]{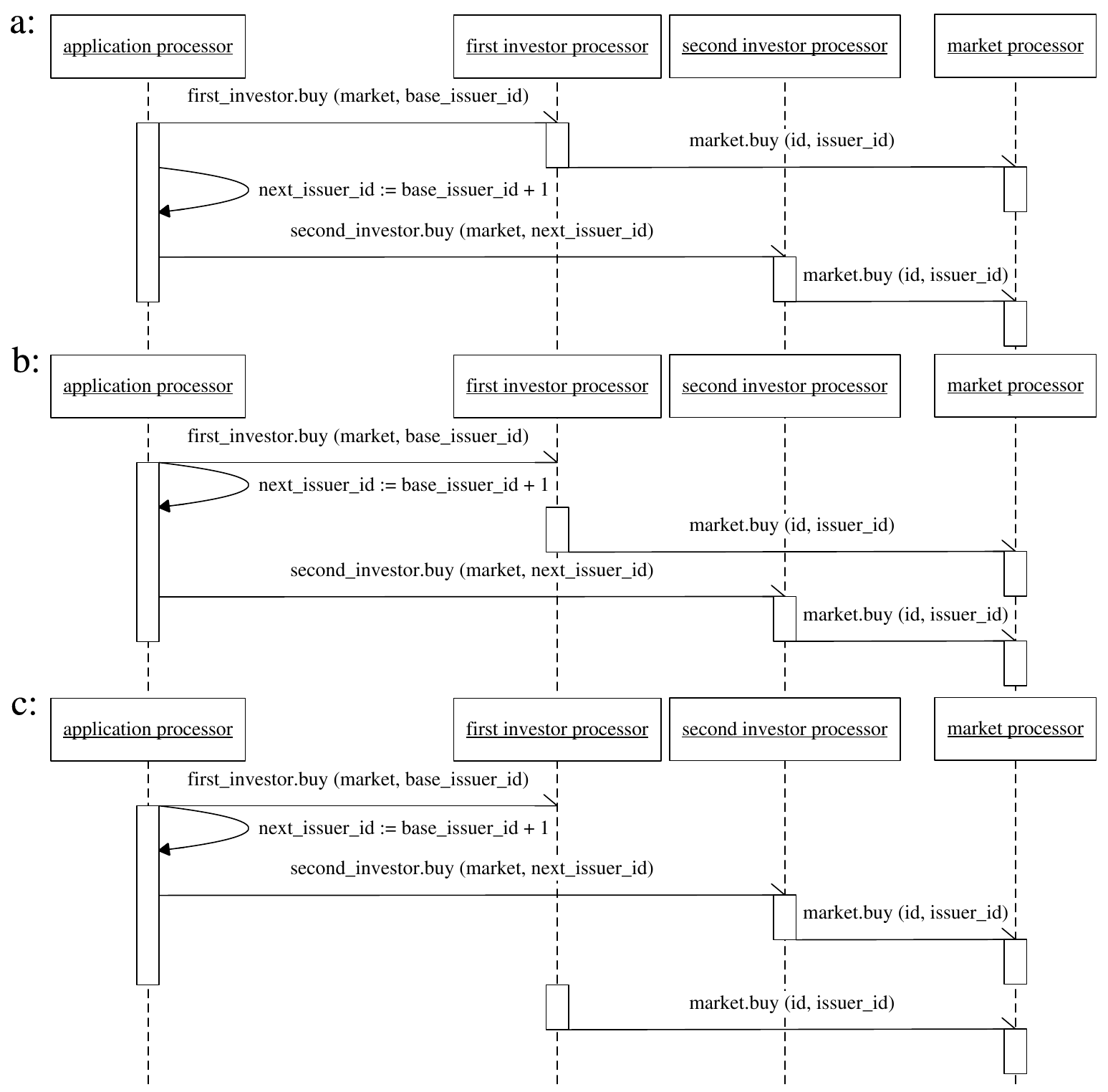}
  \caption{Three possible physical processor schedules for the market example}
  \label{fig:market example possible physical processor schedules}
\end{figure}

\figurereference{fig:market example possible physical processor schedules} depicts a number of possible physical processor schedules for this example. The difference between schedules $a$ and $b$ is that in $a$, the application sets the local variable \lstinline[language=SCOOP]!next_issuer_id! after the first investor buys its share from the market, whereas in $b$ the variable is set before this event. In schedule $c$, the second investor buys its share before the first investor does.
Schedules $a$ and $b$ give rise to the same behavior on the market, whereas schedule $c$ causes the transaction to be reversed: the second investors gets to buy its share first. The reason is that changes in the update of local variables do not influence shared objects, whereas the order of \emph{critical events} does. In SCOOP, the only critical events occur in the synchronization step, i.e., when the scheduler approves a locking request. We regard two physical processor schedules as equivalent if they have the same order of locking requests. A \emph{logical processor schedule} denotes an equivalence class of physical processor schedules, i.e., physical processor schedules where the scheduler approves the locking requests in the same order. \sectionreference{sec:recording logical processor schedules} describes the implementation of logical processor schedules.

\subsection{Recording logical processor schedules}\label{sec:recording logical processor schedules}
A logical processor schedule consists of one interval list per processor. An \emph{interval list} is a sequence of intervals that keeps track of a processor's approved locking requests. The scheduler uses a \emph{global counter} with value $counter_{g}$ to number the approved locking requests. An \emph{interval} $[l, u]$ is defined by a lower global counter value $l$ and an upper global counter value $u$, such that the locking requests with numbers in $[l, u]$ belong to the same processor and no locking request with a number in an adjacent interval belongs to the same processor.

Once the recorder is activated, the scheduler executes \algorithmreference{alg:record}. To detect when a new interval should start, the scheduler maintains for each processor a \emph{local counter} with value $counter_{l}$ and a \emph{local counter base} with value $base_{l}$. The local counter base of a processor $p$ stores the value of the global counter at the point where the scheduler started recording an interval for $p$. The local counter counts $p$'s locking requests that got approved from the moment where the scheduler started recording the interval for $p$. Processor $p$'s current interval is then given as $[base_{l}[p] + 1, base_{l}[p] + counter_{l}[p]]$.

\begin{algorithm}[!ht]
\eventhandler{\textlangle Initialize\textrangle}{\tcp*[h]{The program starts.}}{
	$counter_{g}$ := $0$; \tcp*[h]{The global counter.}\\
	\ForAll{$p \in processors$}{
		$counter_{l}[p]$ := $\undefinedvalue$; \tcp*[h]{The local counters.}\\
		$base_{l}[p]$ := $\undefinedvalue$; \tcp*[h]{The local counter bases.}\\
		$intervals[p]$ := $()$; \tcp*[h]{The interval lists.}\\
	}
}

\eventhandler{\textlangle Approved \textbar \thinspace $p$\textrangle}{\tcp*[h]{The scheduler approved $p$'s request.}}{
	\uIf{$counter_{l}[p] = \undefinedvalue$}{
			$base_{l}[p]$ := $counter_{g}$\;
			$counter_{l}[p]$ := $1$\;
			$counter_{g}$ := $counter_{g} + 1$\;
		}
	\uElseIf{$counter_{l}[p] \neq \undefinedvalue \wedge counter_{g} = base_{l}[p] + counter_{l}[p]$}{
			$counter_{l}[p]$ := $counter_{l}[p] + 1$\;
			$counter_{g}$ := $counter_{g} + 1$\;
		}
	\ElseIf{$counter_{l}[p] \neq \undefinedvalue \wedge counter_{g} \neq base_{l}[p] + counter_{l}[p]$}{
			$intervals[p]$ := $intervals[p] \bullet [base_{l}[p] + 1, base_{l}[p] + counter_{l}[p]]$\;
			$base_{l}[p]$ := $counter_{g}$\;
			$counter_{l}[p]$ := $1$\;
			$counter_{g}$ := $counter_{g} + 1$\;
		}	
}

\eventhandler{\textlangle Terminate\textrangle}{\tcp*[h]{The program terminates.}}{
	\ForAll{$p \in processors$}{
		\If{$counter_{l}[p] \neq \undefinedvalue$}{
			$intervals[p]$ := $intervals[p] \bullet [base_{l}[p] + 1, base_{l}[p] + counter_{l}[p]]$\;
			write ($p$, $intervals[p]$)\;
		}
	}
}
\caption{Record\label{alg:record}}
\end{algorithm}

Whenever the scheduler approves a locking request $r$ of a processor $p$, it goes through the following checks. If $p$'s local counter is undefined, then $p$ does not have an interval yet, and thus $r$ belongs to a new interval for $p$. Hence, the scheduler starts recording a new interval for $p$. 

If $p$'s local counter is defined and $counter_{g} = base_{l}[p] + counter_{l}[p]$, then the scheduler is currently recording an interval for $p$, and $r$ belongs to this interval. This can be seen as follows. If the scheduler would have approved locking requests of any other processor $q$ since it started recording $p$'s interval, then the scheduler would have incremented the global counter, but not $p$'s local counter. Thus the equation would not hold. Hence, the scheduler did not approve locking request of other processors and thus $r$ belongs to $p$'s current interval. 

If $p$'s local counter is defined and $counter_{global} \neq base_{local} + counter_{local}$, then the scheduler is currently recording an interval for $p$, and $r$ belongs to a new interval. This can be seen as follows. If the scheduler would not have approved locking requests of any other processor $q$, since it started recording $p$'s current interval, then only $p$ would have incremented the global counter and its local counter. Thus the equation would hold. Hence, the scheduler must have approved one or more locking requests of other processors and thus $r$ belongs a new interval on $p$. In this case, the scheduler finishes $p$'s current interval and adds $r$ to a new interval. 

At the end of the program execution, the scheduler checks for each processor whether there is any pending interval, in which case it adds the interval to the respective interval list.

\begin{figure}[ht]
  \centering
  \includegraphics[width=\textwidth]{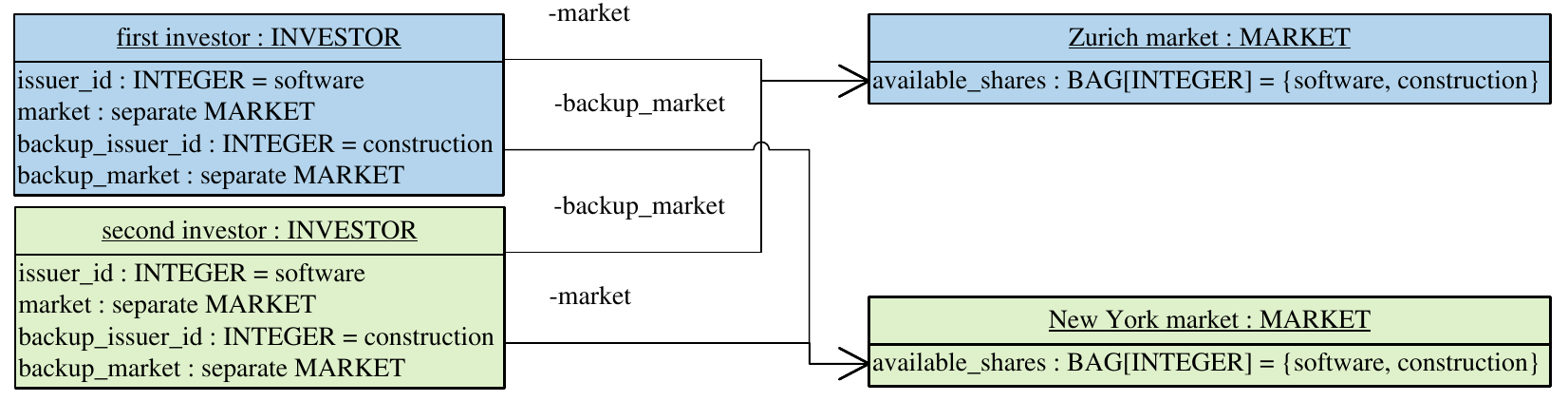}
  \caption{Object structure for the market example}
  \label{fig:market example object structure}
\end{figure}

\begin{figure}[!ht]
  \centering
  \includegraphics[width=\textwidth]{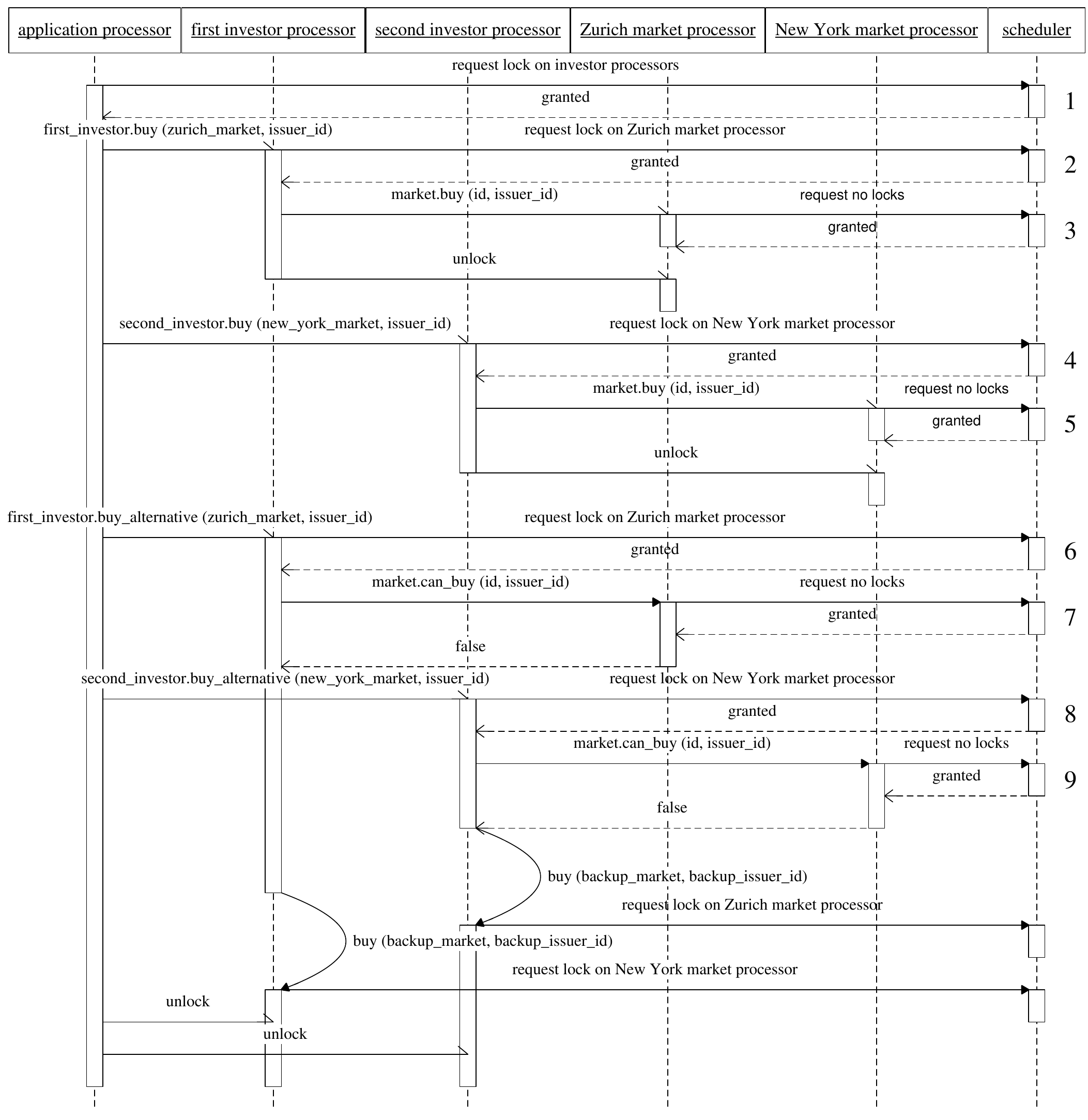}
  \caption{A physical processor schedule of the market example in detail. The numbers next to the scheduler lifeline indicate the approved locking requests.}
  \label{fig:market example physical processor schedule}
  \vspace*{1cm}
\end{figure}

Consider again the market example. Assume the investor class has an additional feature \lstinline[language=SCOOP]!buy_alternative!, which allows an investor to buy a share if possible; if it is not possible, a backup share is bought. For this reason, each investor has a backup market and an identifier of a backup issuer.
\begin{lstlisting}[language=SCOOP]
buy_alternative (market: separate MARKET; issuer_id: INTEGER)
		-- Try to buy a share of the issuer on the market.
		-- If this fails, buy some backup share on the backup market.
	do
		if market.can_buy (id, issuer_id) then
			market.buy (id, issuer_id)
		else
			buy (backup_market, backup_issuer_id)
		end
	end
\end{lstlisting}
Consider the setup in \figurereference{fig:market example object structure}.
%
Assume that a new transaction asks each investor to buy at least one share of the software company by calling \lstinline[language=SCOOP]!buy! and then \lstinline[language=SCOOP]!buy_alternative!. The schedule in \figurereference{fig:market example physical processor schedule} leads to a deadlock because the two investors hold a lock on one market while trying to lock the other market; however, not all possible schedules exhibit the problem.
%
The proposed technique produces the following logical processor schedule: application: $[1, 1]$, first investor: $[2, 2] \bullet [6, 6]$, second investor: $[4, 4] \bullet [8, 8]$, Zurich market: $[3, 3] \bullet [7, 7]$, and New York market: $[5, 5] \bullet [9, 9]$. \sectionreference{sec:replaying logical processor schedules} shows how to replay this logical processor schedule to reproduce the deadlock.
%
%

\subsection{Replaying logical processor schedules}\label{sec:replaying logical processor schedules}
To replay a logical processor schedule, the scheduler once again uses a global counter $counter_{g}$; this time the global counter represents the number of the locking request that the scheduler wants to approve next. To replay, the scheduler executes \algorithmreference{alg:replay}.

\begin{algorithm}[!ht]
\eventhandler{\textlangle Initialize\textrangle}{\tcp*[h]{The program starts.}}{
	$counter_{g}$ := $1$; \tcp*[h]{The global counter.}\\
	\ForAll{$p \in processors$}{
		$intervals[p]$ := read ($p$); \tcp*[h]{The interval lists.}\\
	}
}

\eventhandler{\textlangle Check \textbar \thinspace $p$\textrangle}{\tcp*[h]{The scheduler checks on $p$'s request.}}{
	\uIf{$\exists [l, u] \in intervals[p] \colon l \leq counter_{g} \leq u$}{
		$counter_{g}$ := $counter_{g} + 1$\;
		\triggerinstruction{\textlangle Ok \textrangle}; \tcp*[h]{The request is next.}\\
	}
	\Else{
		\triggerinstruction{\textlangle NotOk \textrangle}; \tcp*[h]{The request is not next.}\\
	}
}
\caption{Replay\label{alg:replay}}
\end{algorithm}

To begin, the scheduler gets ready to approve the first locking request.
Whenever the scheduler is about to approve a locking request $l$ of a processor $p$, the scheduler first checks whether $l$ is next. To do so, the scheduler consults $p$'s interval list and checks whether it contains an interval with $counter_{g}$. If the interval list contains such an interval, then the scheduler approves the locking request and gets ready to approve the next locking request, i.e., it increments the global counter. If the interval list does not contain such an interval then the scheduler tries another locking request.

To replay the logical processor schedule from \sectionreference{sec:recording logical processor schedules}, the scheduler initializes the global counter to $1$. As soon as the application sends a locking request, the scheduler approves and increments the global counter to $2$. The first two calls on the investors cause them to each send a locking request. The scheduler lets the first investor proceed and sets the global counter to $3$. The second investor must wait because its interval list does not contain the current global counter value. The first investor calls the Zurich market, whose locking request the scheduler approves right away. Now the global counter is at $4$, and the scheduler lets the second investor and the New York market proceed. As a consequence, the global counter reaches $6$. In the meantime, the application performed two more calls to the investors. In sequence, the scheduler approves the locking requests of the first investor, the Zurich market, the second investor, and the New York market. The deadlock is guaranteed.

%% file: conclusion.tex
\section{Conclusion}\label{sec:conclusion}
While the SCOOP model protects developers from introducing data races, its run-time system is complex; this makes errors such as deadlock hard to analyze without the ability to reproduce them. We introduced a record-replay technique to record and reproduce the execution of SCOOP programs. The technique uses the idea of logical thread schedules~\cite{choi-srinivasan:1998:replay_for_concurrent_programs} to abstract from non-critical events. The simplicity of the SCOOP model helped to apply this technique: the approvals of locking request are the only relevant critical events.

The ability to replay executions using logical processor schedules is an important component to test SCOOP programs. In future work, schedules may be generated in order to drive programs systematically through different orders.

\paragraph{Acknowledgments}
We thank Andrey Nikonov and Andrey Rusakov who contributed during a research internship. This work is part of the SCOOP project at ETH Zurich, which has benefited from grants from the Hasler Foundation, the Swiss National Foundation, Microsoft (Multicore award), and ETH (ETHIIRA).